LETTER

# Establishing personal trust-based connections in distributed teams

Fabio Calefato[1] | Filippo Lanubile[2]

[1]Dipartimento Jonico, University of Bari, Bari, Italy

[2]Dipartimento di Informatica, University of Bari, Bari, Italy

**Correspondence**
Filippo Lanubile, Dipartimento di Informatica, University of Bari, Via Orabona, 4 – Dip. Informatica, 70125 Bari, Italy.
Email: filippo.lanubile@uniba.it

Trust is a factor that dramatically contributes to the success or failure of distributed software teams. We present a research model showing that social communication between distant developers enables the affective appraisal of trustworthiness even from a distance, thus increasing project performance. To overcome the limitations of self-reported data, typically questionnaires, we focus on software projects following a pull request-based development model and approximate the overall performance of a software project with the history of successful collaborations occurring between developers.

## 1 | INTRODUCTION

Computer-Supported Cooperative Work (CSCW) addresses collaboration problems faced by virtual teams, including communication breakdowns, coordination problems, and lack of knowledge about colleagues' activities.[1] Distributed software teams are no different. When developers work for large-scale, distributed projects, "distances" (ie, temporal, geographical, and socio-cultural) get in the way, aggravating collaboration problems between teams. Software development is, in fact, an intensely collaborative process where people are required to constantly interact with others to create, share, and integrate information.

In their study on the challenges of distributed development caused by distance, Agerfalk and Fitzgerald[2] stressed the negative impact of *reduced trust* faced by distributed teams. Trust within teams typically grows and reinforces through direct contact. Face-to-face communication is the most effective solution for team members to the problem of creating ties with others and becoming aware of both technical aspects, such as terminology and established procedures, and even more subtle aspects, such as existing social connections between team members, norms, and cultural differences. Despite being so vital for establishing connections between developers, face-to-face interaction is the activity that is mostly affected by distance in distributed development contexts. Over the last 2 decades, software development organizations have become more and more distributed and, yet, the following research question remains mostly unanswered: *How do we establish personal, trust-based connections between members of distributed teams*?

## 2 | BACKGROUND

The topic of trust has received a considerable amount of attention in several research domains, from cognitive science to economy and software engineering. To date, several definitions of trust have been given due to the complexity of the matter, which involves both interpersonal relationships (eg, cultural issues between trustee and trustor) and facets of human behavior (eg, personal traits).

Trust may be defined as the belief that the trustee will behave as expected by the trustor.[3] This definition explicates that the trustor decides to take the risk of trusting the trustee based on the judgment of existing information available from both direct and indirect observation. As such, trust is often modeled as of a 3-phase process that involves *formation*, *dissolution*, and

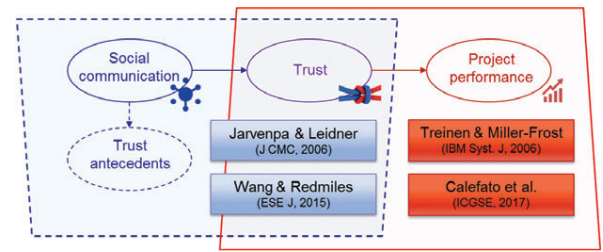

**FIGURE 1** The proposed research framework to investigate how increasing social communication and trust can turn into increased project performance

*restoration*, depending on whether the expectations of the trustor are met.[4] In other words, positive trust emerges when others' actions meet our expectation; otherwise, negative trust (or mistrust) arises.

Other definitions of trust distinguish between cognitive (or rational) and affective (or social) perspectives.[5] Accordingly, cognitive trust is defined as the expectations about others' skills and reliability when executing critical tasks that the trustor cannot monitor. Conversely, affective trust relates to reciprocal emotional ties, concerns, and care between the trustee and the trustor. When affective trust is established, the trustee is pushed to do something for the trustor because the former feels responsible and perceives the action to take as a moral duty.

Albeit important to teams of any kind, trust is vital to the well-being of distributed software teams as it prevents physical distance from leading to psychological distance too.[6] Indeed, reduced trust exacerbates the feeling of being distinct teams with different and conflicting goals, affects benevolence toward others when settling differences, and decreases team members' disposition to share information and cooperate in solving issues.[7,8]

## 3 | RESEARCH MODEL

The existing literature has proposed several approaches for fostering trust.[9] These models build on the underlying idea that trust develops along several dimensions called *trust antecedents*,[10] that is, the properties that trigger the trustor's judgment when evaluating the trustworthiness of the trustee.

Jarvenpaa and Leidner[11] studied how trust evolves in global teams who only had chances of interaction through computer-mediated communication. They observed that teams who ended up establishing a low level of trust were those lacking in social communication at the beginning of projects. Instead, those teams who had an initial focus on social communication, which later reduced in favor of technical and task-focused interaction, were capable of establishing a high level of trust by the end of projects. Consistently, Wang and Redmiles[12] studied the communication patterns of globally distributed software developers interacting over Internet Relay Chat (IRC) and found that positive, nonwork-related interactions promoted trust and cooperation within teams. Accordingly, in Figure 1 (see the blue box on the left), we depict a research model showing that social communication between distant developers helps to get access to trust antecedents, otherwise unavailable in distributed contexts, which enable the affective appraisal of trustworthiness even from a distance.

One common limitation of prior research is that there is no explicit measure of *how much* improving trust contributes to project performance. Indeed, it is challenging to prove the existence of a cause and effect relations between the amount of trust established between developers and the performance of a software project as many confounding factors that are difficult or impossible to control (eg, project type, individual skills and experience, and company's culture) may interfere. Treinen and Miller-Frost[13] studied how trust develops across remote sites of distributed software projects at IBM. They observed that distant sites with increased level of trust could resolve issues from afar (eg, through videoconferences) and even anticipate future difficulties, thus resulting in increased overall project efficiency. However, to the best of our knowledge, no previous study has provided evidence that directly connects and quantified the relation existing between the amount of trust established between developers and project performance.

Besides, existing research on trust has so far relied mostly on self-reported data (ie, questionnaires) to measure trust levels.[9] In our previous work,[14] we proposed to overcome this limitation by approximating the overall performance of a software project (intended as broadly as possible) with the history of successful collaborations between developers, which constitute a large part of the project lifecycle. We called a *successful collaboration* any situation involving at least 2 software developers who cooperate successfully (eg, adding a new feature, refactoring poor-quality code, and fixing a bug), thus producing a project advancement. A successful collaboration is, therefore, a finer-grain unit of analysis that enables measuring the extent to which trust increases project performance more directly. Furthermore, recent work has found that the *social distance* between developers (eg, history of previous interactions and the existence of connections in social coding platforms) is strongly predictive of whether a contribution will be accepted.[15] Many open source software projects rely today on GitHub, a social-coding platform that enables developers to collaborate through pull requests. A *pull request* is a proposed set of source code changes submit-ted by a potential contributor that is integrated upon inspection and acceptance by one or more core team members. As such, according to our definition, an accepted pull request represents a successful collaboration between the pull-request contributor and the integration manager(s). Hence, we argue that trust catalyzes the process of reviewing and accepting pull requests, which, in turn, boost project performance (see the red box on the left in Figure 1).

TABLE 1  Breakdown of datasets and results from the Apache Groovy and Drill projects

| Analysis time window | | Apache Groovy March 2015–December 2016 | Apache Drill November 2014–September 2017 |
| --- | --- | --- | --- |
| Core team | | 6 | 22 |
| Overall contributors | | 211 | 88 |
| Closed PRs | Merged | 200 | 668 |
| | Rejected | 18 | 66 |
| Emails | | 4948 | 29 514 |
| Effect of high propensity to trust on merging pull requests | | +34% | +26% |

## 4 | DISCUSSION AND CONCLUSIONS

As a first step, to assess the hypothesis that social communication may foster trust building in distributed software teams, we developed SocialCDE,[16–18] a tool that extends Microsoft Team Foundation Server and GitHub by disclosing information collected from popular social networks such as Facebook, Twitter, and LinkedIn.

In our recent work,[14] we took one further step toward collecting quantitative evidence that establishing trust between developers contributes to project performance. Because trust has several facets, we focused initially on the *propensity to trust*, that is, the natural, personal disposition to perceive the others as trustworthy.[19] We run an experiment where we used the IBM Watson Tone Analyzer service to measure the propensity to trust of Apache Groovy project's core team members through the analysis of their emails (see the first column in Table 1). Tone Analyzer has been built upon the Big-five personality model,[20] a general taxonomy of personality traits including openness, conscientiousness, extraversion, agreeableness, and neuroticism. We focused on *agreeableness*, the personality trait associated with the tendency to trust others. In the experiment, we first identified the integration managers from the Groovy projects; then, we parsed the entire email archive to compute their agreeableness score and use it as a proxy measure of their *high* vs *low* level of propensity to trust. Finally, we built a logistic regression model to estimate the probability of pull requests being accepted after being reviewed by the project's integration managers. As compared to those with a low level of propensity to trust, we found that the distributed team members with a high propensity to trust are more likely (+34%) to accept external contributions in form of pull requests.

To increase the validity of our findings, here we report the results from the analysis of Drill, another project from the Apache Software Foundation. In particular, we identified 22 core team members out of 88 overall contributors. As of this writing, the project counts almost 800 closed pull requests (ie, merged or rejected). Again, we mined the project's mailing lists archives to compute the agreeableness score for each of the core team members. In line with our previous findings, we found that pull requests reviewed by Drill integrators who have a high propensity to trust get accepted more often (+26%).

To the best of our knowledge, this is the first attempt at quantifying the effects that trust or other developers' personal traits have on software projects that follow a pull request-based development model. As such, the main result of our work is the initial evidence that the personality traits of the integrators who perform the review of a code contribution are a strong predictor of the probability of its merge with the code base. This is a novel finding that highlights the role of trust in the execution of complex, software development tasks such as code review.

As future work, first, we intend to replicate the experiment to compare several tools and assess their reliability in extracting personality from different types of technical text other than emails (eg, Q&A posts, commits messages, and error reports). Second, we intend to enlarge the dataset in terms of number and types of projects to understand whether developers' trust changes depending on the project, how it evolves over time, and whether negative events (eg, missed deadlines) recorded in project lifecycles can be traced back to observed cases of trust dissolution and restoration involving core team members.

ORCID

*Fabio Calefato*  http://orcid.org/0000-0003-2654-1588
*Filippo Lanubile*  http://orcid.org/0000-0003-3373-7589